# Assessment of Intra-channel Fiber Nonlinearity Compensation in 200 GBaud and Beyond Coherent Optical Transmission Systems

Zhiyuan Yang, Mengfan Fu, Yihao Zhang, Qizhi Qiu, Lilin Yi, Weisheng Hu, and Qunbi Zhuge, *Senior Member*, *IEEE*, *Senior Member*, *Optica*

*Abstract*—In this paper, we investigate and assess the performance of intra-channel nonlinearity compensation (IC-NLC) in long-haul coherent optical transmission systems with a symbol rate of 200 GBaud and beyond. We first evaluate the potential gain of ideal IC-NLC in 4 THz systems by estimating the proportion of self-channel interference (SCI) using the split-step Fourier method (SSFM) based simulation with either lumped amplification or distributed amplification. As the symbol rate increases to 300 GBaud, the SCI proportion exceeds 65%. On the other hand, the non-deterministic polarization mode dispersion (PMD) will impact the effectiveness of IC-NLC, especially for ultra-high symbol rate systems. Therefore, we investigate the power spectral density of the residual nonlinear noise after ideal IC-NLC in the presence of PMD. The results indicate that the gain of ideal digital backpropagation (IDBP) decreases by 3.85 dB in 300 GBaud erbium-doped fiber amplifier (EDFA)-amplified links with a PMD parameter of 0.05 ps/km$^{1/2}$, and 5.09 dB in distributed Raman amplifier (DRA)-amplified links. Finally, we evaluate the potential gains of practical IC-NLC in C-band wavelength-division multiplexing (WDM) systems by employing the low-pass-filter assisted digital backpropagation (LDBP). As the symbol rate increases from 100 GBaud to 300 GBaud, the gain of 20-step-per-span (20-stps) LDBP increases from 0.53 dB to 0.87 dB for EDFA-amplified links, and from 0.89 dB to 1.30 dB for DRA-amplified links. Our quantitative results show that for 200 GBaud and beyond systems, there is a sizable gain to achieve by compensating for intra-channel nonlinearity even with a large non-deterministic PMD.

## I. Introduction

Driven by the emergence of new technologies such as cloud computing, Internet of Things, machine learning and so forth, the global demand for optical communication capacity is experiencing exponential growth [1]. Improving the symbol rate of an optical signal is an efficient approach to satisfy this ever-increasing demand. Over the past decade, industry efforts have been dedicated to increasing the symbol rate from ~35 GBaud to ~200 GBaud [2]. In the future, the symbol rate is expected to approach 300 GBaud or even higher. In the meantime, fiber nonlinearity remains a fundamental limiting factor to further increase the fiber capacity and achievable transmission distance of ultra-high symbol rate optical communication systems [3], [4]. To address the impairment of fiber nonlinearity, various compensation techniques have been developed, encompassing both optical and digital schemes [1], [2]. These include optical phase conjugation (OPC), digital intra-channel nonlinearity compensation (IC-NLC), nonlinear probabilistic shaping and so forth [2], [4], [5]. Of these approaches, IC-NLC has emerged as a promising solution for next generation ultra-high symbol rate systems [4]. On the other hand, the increase of symbol rate will provide an additional potential gain for IC-NLC algorithms. Therefore, it is valuable to evaluate the performance of IC-NLC in the design of ultra-high symbol rate optical communication systems.

In wavelength-division multiplexing (WDM) systems, fiber nonlinearity can be classified into three categories: self-channel interference (SCI, which is also called intra-channel nonlinearity), cross-channel interference (XCI), and multi-channel interference (MCI) [6], [7]. In the past decade, various IC-NLC algorithms which can compensate for SCI have been widely investigated, such as digital backpropagation (DBP) [8]-[12], perturbative nonlinearity compensation (PNC) [13]-[16] Volterra series [17], [18], and machine learning methods [19], [20]. In a fixed bandwidth system where the total amount of fiber nonlinear noise remains relatively constant, the potential gain of IC-NLC is positively correlated with the proportion of SCI. Therefore, IC-NLC may achieve a large compensation gain in ultra-high symbol rate systems.

On the other hand, the effectiveness of IC-NLC may be impaired by non-deterministic effects which include nonlinear signal-amplified spontaneous emission (ASE) interaction [21],

Manuscript received 9 June 2024; revised 9 June 2024 and 9 June 2024; accepted 9 June 2024. Date of publication 9 June 2024; date of current version 9 June 2024. This work was supported in part by the Shanghai Pilot Program for Basic Research-Shanghai Jiao Tong University under Grant 21TQ1400213, and in part by the National Natural Science Foundation of China under Grant 62175145. (*Corresponding author: Qunbi Zhuge.*)

Zhiyuan Yang, Mengfan Fu, Yihao Zhang, Qizhi Qiu, Lilin Yi, Weisheng Hu and Qunbi Zhuge are with the State Key Laboratory of Advanced Optical Communication Systems and Networks, Department of Electronic Engineering, Shanghai Jiao Tong University, Shanghai 200240, China (e-mail: zhiyuan.yang@sjtu.edu.cn; mengfan.fu@sjtu.edu.cn; yihao.zhang@sjtu.edu.cn; qizhi.qiu@sjtu.edu.cn; lilinyi@sjtu.edu.cn; wshu@sjtu.edu.cn; qunbi.zhuge@sjtu.edu.cn).

Color versions of one or more figures in this article are available at https://doi.org/10.1109/JLT.2022.3181340.

Digital Object Identifier 10.1109/JLT.2022.3181340.

[22] and stochastic polarization dependent nonlinearity interaction [23], [24]. The non-deterministic distributed polarization mode dispersion (PMD) impairs the effectiveness of IC-NLC more significantly than the nonlinear signal-ASE interaction [23]. The non-deterministic PMD emerges as a critical factor influencing IC-NLC performance. Since the refractive indexes of the fast and slow principal states of the birefringence vary with wavelength, the correlation of PMDs at different frequencies decreases as the frequency difference increases [25]. Therefore, the influence of non-deterministic PMD on the effectiveness of IC-NLC may become more pronounced in ultra-high symbol rate systems.

To sum up, a thorough quantitative analysis of how much the SCI proportion increases with symbol rate and how much nonlinear compensation gain can be achieved for next generation coherent systems remains a very important question. The assessment will provide guidance to both academia and industry in conducting future research and design on fiber nonlinearity compensation for next generation ultra-high symbol rate systems.

In this paper, we quantitatively evaluate the gain of IC-NLC in systems with a symbol rate of 200 GBaud and beyond. First, we provide a comprehensive quantitative assessment of SCI proportion in systems with different symbol rates. The split-step Fourier method (SSFM) based simulation is adopted to calculate the SCI proportion. For C-band standard single mode fiber (SSMF) systems with erbium-doped fiber amplifier (EDFA) amplification, the SCI proportion increases from 45.4% to 66.5%, as the symbol rate increases from 100 GBaud to 300 GBaud. As the SCI becomes the majority of fiber nonlinearity, the potential gain of IC-NLC becomes much larger. Second, we thoroughly evaluate the influence of non-deterministic PMD on ideal IC-NLC through theoretical and numerical studies. We extend the closed-form expressions of the power spectral density of the residual nonlinear noise induced by PMD to systems with distributed Raman amplifier (DRA) amplification, based on the expressions in systems with EDFA amplification [23]. These expressions are used to evaluate the influence of PMD on the effectiveness of IC-NLC. The results show that the influence of non-deterministic PMD on the effectiveness of IC-NLC becomes more pronounced in ultra-high symbol rate systems. In EDFA/DRA-amplified links with a PMD parameter of 0.05 ps/km$^{1/2}$ and a symbol rate of 300 GBaud, the interaction between PMD and nonlinearity causes a 3.85 / 5.09 dB decrease in the gain of ideal digital backpropagation (IDBP), respectively. Finally, the potential gain of practical IC-NLC considering the complexity constraint is evaluated to provide a reference for future research on fiber nonlinearity compensation in next generation ultra-high baud systems. Taking the low-pass-filter assisted DBP (LDBP) [10] as an example, as the symbol rate increases from 100 GBaud to 300 GBaud, the gain of 20-step-per-span (20-stps) LDBP increases from 0.53 dB to 0.87 dB for EDFA-amplified links, and from 0.89 dB to 1.30 dB for DRA-amplified links.

The remainder of this paper is organized as follows. In Section II, the SCI proportion is first evaluated with varying symbol rates. Different fiber types and amplification methods are considered for comparison. Then the theoretical analysis of the impact of PMD is described, and the formulas to calculate the power spectral density of the residual nonlinear noise induced by PMD are provided. In Section III, the accuracy of the power spectral density formulas and the influence of the residual nonlinear noise are investigated by simulations. In Section IV, the performance of LDBP in C-band WDM systems is discussed. The performance of IDBP is also assessed for comparison. Finally, the conclusions are drawn in Section V.

TABLE I
SIMULATION PARAMETERS

| Parameters | Values |
|---|---|
| Symbol Rate (GBaud) | 100/200/300 |
| Channel Spacing (GHz) | 110/220/330 |
| Total Bandwidth | C-band (~4 THz) |
| Number of Spans | 10 |
| Span Length (km) | 80 |
| EDFA Gain (dB) | 16 |
| DRA On-off Gain (dB) | ~16 |
| DRA Pump Type | Backward |

TABLE II
FIBER PARAMETERS

| Fiber | α (dB/km) | D (ps·nm$^{-1}$·km$^{-1}$) | γ (W$^{-1}$·km$^{-1}$) |
|---|---|---|---|
| SSMF | 0.2 | 16.7 | 1.3 |
| PSCF | 0.17 | 20.1 | 0.8 |
| NZDSF | 0.22 | 3.8 | 1.5 |
| G.654E | 0.17 | 22 | 0.9 |

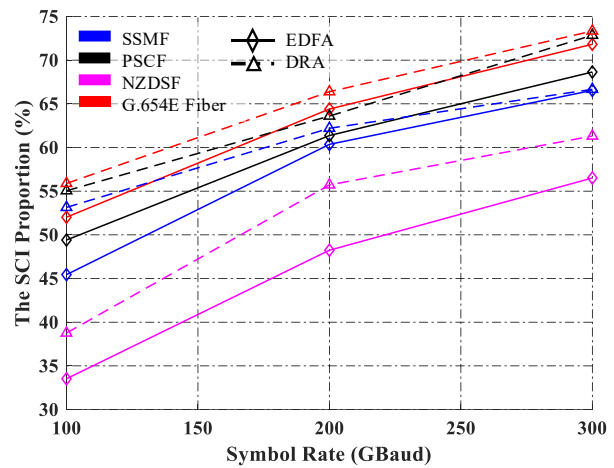

Fig. 1. The SCI proportion vs. the symbol rate in C-band WDM systems.

## II. THEORETICAL ANALYSIS OF IC-NLC'S GAIN

In this section, a theoretical analysis is provided to evaluate the potential gain of IC-NLC. We first assess the achievable gain of IC-NLC by estimating the proportion of SCI within fiber nonlinearity. Subsequently, we study the impact of non-deterministic distributed PMD and provide the analytical expressions of the residual nonlinear noise induced by PMD in ideal IC-NLC compensated systems with DRA amplification.

### A. The Increase of SCI Proportion

In systems employing ideal IC-NLC, SCI can be completely eliminated. Consequently, the proportion of SCI determines the potential of IC-NLC. While the correlation between the SCI proportion and symbol rate has been investigated, a quantitative assessment of how the SCI proportion increases with increasing symbol rates remains a very important question. In order to estimate the SCI proportion quantitatively, the SSFM-based simulation is conducted to calculate the power of nonlinear interference (NLI). The simulation parameters are shown in Table I. The detailed configuration in our evaluation is described as follows. The symbol rate varies from 100 GBaud to 300 GBaud with a step of 100 GBaud, while the channel spacing varies from 110 GHz to 330 GHz with a step of 110 GHz. For each symbol rate, we conduct single-channel and multi-channel simulations. During single-channel SSFM propagation, we only simulate fiber impairments, while excluding ASE noise and other noise sources. This simulation methodology ensures that the received noise contains only SCI. To reasonably investigate the total power of fiber nonlinearity, the number of channels is a free set-up parameter in multi-channel simulations, whereas the total bandwidth is kept approximately 4 THz. Similarly, in multi-channel simulations, we maintain the same simulation setups as in single-channel simulations, thereby ensuring that the received signal is only affected by fiber nonlinearity noise. The transmission systems consist of identical spans and the loss of each span is compensated for by optical amplifiers completely [7]. Four different types of optical fibers are considered, and their parameters are shown in Table II [7], [26]. Lumped amplification employing EDFAs and distributed amplification employing backward DRAs are both considered.

Simulation results are shown in Fig. 1. As the symbol rate increases, there is a notable increase in the SCI proportion. Taking the SSMF system with EDFA amplification as an example, when the symbol rate increases from 100 GBaud to 300 GBaud, the SCI proportion of the center channel increases from 45.4% to 66.5%. The SCI proportion of pure-silica-core fiber (PSCF) and G.654E fiber is similar to that of SSMF. On the other hand, compared to SSMF, PSCF and G.654E, the SCI proportion of non-zero dispersion shifted fiber (NZDSF) is much lower. This phenomenon is attributed to the smaller dispersion-induced walk-off of signals during transmission. For systems with DRA amplification, the results are quite similar to those in systems with EDFA amplification.

In summary, the SCI proportion increases as the symbol rate increases in systems with different fiber types and amplification schemes. In 4 THz SSMF systems with a symbol rate of 300 GBaud, the SCI proportion is beyond 65%. Therefore, IC-NLC is anticipated to achieve high compensation gains by compensating for SCI in ultra-high symbol rate systems.

*B. The Impact of Non-deterministic PMD*

In this subsection, we focus on the impact of non-deterministic PMD which will impair the effectiveness of IC-NLC. We first arrive at the closed-form expressions of the power spectral density of the residual nonlinear noise induced by PMD in EDFA-amplified systems based on perturbation theory. Subsequently, we extend the power spectral density expressions of the PMD-induced residual nonlinear noise to DRA-amplified systems.

*1) Systems with EDFA amplification*

To simulate signal propagation in optical fibers, the Manakov equation can be used [27]:

$$\frac{\partial \boldsymbol{E}(t,z)}{\partial z} + \frac{j\beta_2}{2}\frac{\partial^2 \boldsymbol{E}(t,z)}{\partial t^2} = \frac{8}{9}j\gamma|\boldsymbol{E}(t,z)|^2\boldsymbol{E}(t,z) - \frac{\alpha}{2}\boldsymbol{E}(t,z) \quad (1)$$

where $\boldsymbol{E}(t,z) = [E_x(t,z) \quad E_y(t,z)]^T$ is the electrical field column vector. $\alpha$ is the fiber attenuation coefficient. $\beta_2$ is the group velocity dispersion coefficient. $\gamma$ is the fiber nonlinear coefficient. To consider polarization effects such as PMD in optical fibers, the fast rotation of the state of polarization (RSOP) can be modeled as a 2×2 unitary matrix in the Cayley-Klein model [28] as

$$R = \begin{bmatrix} cos\varphi cos\kappa + jsin\varphi sin\kappa & sin\varphi cos\kappa - jcos\varphi sin\kappa \\ -sin\varphi cos\kappa - jcos\varphi sin\kappa & cos\varphi cos\kappa - jsin\varphi sin\kappa \end{bmatrix} \quad (2)$$

where $\varphi$ and $\kappa$ are the 2 real values. Based on the simplified polarization model, the PMD can be combined with RSOP and calculated as

$$R^{-1}\begin{bmatrix} e^{j\omega\Delta\tau/2} & 0 \\ 0 & e^{-j\omega\Delta\tau/2} \end{bmatrix} R \quad (3)$$

where the delay $\Delta\tau$ in the arrival of the polarized components is called the differential group delay (DGD), which is given by

$$\Delta\tau = \sqrt{\frac{3\pi}{8}} D_p \sqrt{L} \quad (4)$$

where $D_p$ is the PMD parameter. $L$ is the fiber length of each rotation of polarization state.

Based on (1), the deterministic fiber nonlinearity can be numerically computed by SSFM. Therefore, it can be completely compensated for by IDBP which realizes the inverse process of SSFM. Based on the perturbation theory [13], the received signal is represented as

$$\boldsymbol{E}(t,z) = \boldsymbol{E}_0(t,z) + \Delta\boldsymbol{E}(t,z) \quad (5)$$

where $\boldsymbol{E}_0(t,z)$ is the solution of linear propagation, and $\Delta\boldsymbol{E}(t,z)$ is the fiber nonlinear perturbation. The Manakov equation for $\Delta\boldsymbol{E}(t,z)$ is as follows:

$$\frac{\partial}{\partial z}\Delta\boldsymbol{E}(t,z) + j\frac{\beta_2}{2}\frac{\partial^2}{\partial t^2}\Delta\boldsymbol{E}(t,z) =$$
$$j\frac{8}{9}\gamma|\boldsymbol{E}_0(t,z)|^2\boldsymbol{E}_0(t,z) - \frac{\alpha}{2}\Delta\boldsymbol{E}(t,z) \quad (6)$$

(6) can be solved in the frequency domain, with the solution given by:

$$\Delta\widetilde{\boldsymbol{E}}(\omega,z) = j\frac{8}{9}\gamma \int_{-\infty}^{\infty} |\boldsymbol{E}_0(t,0)|^2 \boldsymbol{E}_0(t,0) e^{-j\omega t} dt$$
$$\times \frac{1-e^{-j\Delta\beta z - \alpha z}}{j\Delta\beta + \alpha} \quad (7)$$

where $\Delta\beta$ represents the group velocity difference.

In a PMD impaired system deploying IDBP, (7) can be rewritten as:

$$\Delta\widetilde{\boldsymbol{E}}^p(\omega,z) = j\frac{8}{9}\gamma \int_{-\infty}^{\infty} |\widetilde{\boldsymbol{E}}_0^p(t,0)|^2 \widetilde{\boldsymbol{E}}_0^p(t,0) e^{-j\omega t} dt$$
$$\times \frac{1-e^{-j\Delta\beta z - \alpha z}}{j\Delta\beta + \alpha} \quad (8)$$

where $\Delta\widetilde{\boldsymbol{E}}^p$ and $\widetilde{\boldsymbol{E}}_0^p$ are the fiber nonlinearity perturbation and the linear propagation solution influenced by PMD, respectively. Therefore, residual intra-channel fiber nonlinearity after IDBP exists due to the stochastically changing polarization states [23], [24], which can be expressed as:

$$\Delta \tilde{E}'_{x/y}(\omega,z) = j\frac{8}{9}\gamma[\Delta \tilde{E}^p(\omega,z) - \Delta \tilde{E}(\omega,z)] \times \frac{1-e^{-j\Delta\beta z-\alpha z}}{j\Delta\beta+\alpha} \quad (9)$$

Since the evolutions of PMD at different frequencies are different, the polarization states of signals with frequency difference beyond the PMD correlation bandwidth will experience different evolutions [25]. Therefore, the PMD will impact the effectiveness of IC-NLC more significantly as the symbol rate increases [23].

For dispersion uncompensated systems, the residual nonlinear noise generated along each span is independent and does not interfere with each other [23]. Therefore, the summation of the residual nonlinear noise power spectral density at each span can be considered as the overall residual nonlinear noise power spectral density. Following the derivations of the statistics of the Jones matrix [29], the expression of the overall residual nonlinear noise power spectral density, denoted as $I_{PMD}$, in systems with EDFA amplification is given as [23]

$$I_{PMD} = \sum_{M=\frac{1}{2}}^{N_s-\frac{1}{2}} I_{PMD}^M =$$

$$\frac{3\gamma^2 I^3}{64\pi\alpha|\beta_2|}\left\{8N_s \ln\left(\frac{B}{B_0}\right) - \sum_{M=\frac{1}{2}}^{N_s-\frac{1}{2}}[3E_1(M)+E_2(M)]\right\} \quad (10)$$

$$E_1(M) = Ei\left(-\frac{3\pi^3 B^2 LMD_p^2}{64}\right) - Ei\left(-\frac{3\pi^3 B_0^2 LMD_p^2}{64}\right) \quad (11)$$

$$E_2(M) = Ei\left(-\frac{7\pi^3 B^2 LMD_p^2}{64}\right) - Ei\left(-\frac{7\pi^3 B_0^2 LMD_p^2}{64}\right) \quad (12)$$

where $\overline{\Delta\tau_m} = \sqrt{LM}D_p$ is the average DGD after $M$ spans. $D_p$ denotes the fiber PMD parameter. $\Delta f$ denotes the frequency difference. $N_s$ is the number of spans. $I$ and $I_{PMD}^M$ denote the signal power spectral density and the $M^{th}$ span's residual nonlinear noise power spectral density, respectively. $B_0$ is expressed as $B_0 = \frac{|\beta_2|}{2\pi^2\alpha B}$, where $B$ denotes channel bandwidth [23]. $Ei(x)$ is the exponential integral function defined as $Ei(x) = \int_{-\infty}^{x} \frac{e^t}{t} dt$.

The performance of systems with ideal IC-NLC is estimated by SNR, which considers both ASE noise, the residual nonlinear noise induced by PMD and inter-channel nonlinearity noise, defined as

$$SNR = \frac{P_{ch}}{P_{ASE}+B\cdot I_{PMD}+P_{NLI}^{inter-channel}} \quad (13)$$

where $P_{ASE}$ and $P_{ch}$ represent the power of ASE noise and signal, respectively. $P_{NLI}^{inter-channel}$ represents the power of inter-channel nonlinearity noise, which is neglected in the computation of SNR within single-channel systems. For multi-channel simulations, $P_{NLI}^{inter-channel}$ is calculated by applying IDBP to completely compensate for SCI in multi-channel systems with a PMD parameter of 0 ps/km$^{1/2}$.

In systems with EDFA amplification, ASE noise power is given by [30]

$$P_{ASE}^{EDFA} = N_s \cdot n_{sp} h \nu_0 (G-1) \Delta \nu_o \quad (14)$$

where $n_{sp}$ is the spontaneous emission factor. $G$ is the gain of EDFA. $h$ is the Planck constant. $\nu_0$ is the signal frequency. $\Delta \nu_o$ is the signal bandwidth.

*2) Systems with DRA Amplification*

In systems with DRA amplification, stimulated Raman scattering (SRS) can be described by ordinary differential equations. The equations that govern the power profile of single pump light and single signal light are (neglecting spontaneous noise) [31], [32]

$$\frac{dP_s(z)}{dz} = \frac{g_R(\nu_s,\nu_p)}{A_{eff}} P_s(z)[P_p^+(z)+P_p^-(z)] - \alpha_s P_s(z) \quad (15)$$

$$\pm\frac{dP_p^\pm(z)}{dz} = -\frac{\nu_p}{\nu_s}\frac{g_R(\nu_s,\nu_p)}{A_{eff}} P_p^\pm(z) P_s(z) - \alpha_p P_p^\pm(z) \quad (16)$$

where $P_s(z)$ and $P_p(z)$ are the signal and pump powers, respectively. $\alpha_s$ and $\alpha_p$ are the fiber loss coefficients for the signal and pump, which are assumed as constants. $\nu_s$ and $\nu_p$ are the signal and pump frequencies, respectively. $g_R(\nu_s,\nu_p)$ is the Raman gain coefficient. $A_{eff}$ is the effective area of fiber. Compared to EDFA-amplified links, DRAs will lead to enhanced nonlinearity effects due to a higher signal power profile across the fiber span in DRA-amplified links [33]. In order to mitigate the nonlinear distortion in DRA-amplified links, backward Raman amplification is generally chosen as the main amplification scheme. Therefore, the following discussion will focus on the backward Raman amplification.

In systems with DRA amplification, the signal power profile along the link needs to be considered to calculate the residual nonlinear noise induced by PMD. To obtain the closed-form solutions of the signal power profile, we assume that pump depletion is negligible and first-order Raman amplification is chosen. Based on these assumptions, the coupled equations (15) and (16) can be solved analytically [34]. The closed-form expression of the normalized backward Raman amplification signal power profile $P_s(z)$ is

$$P_s(z) = \frac{\exp\left[\frac{g_R(\nu_s,\nu_p)}{\alpha_p A_{eff}} P_p(L)\exp(-\alpha_p L)\exp(\alpha_p z)\right]\exp(-\alpha_s z)}{\exp\left(\frac{g_R(\nu_s,\nu_p)}{\alpha_p A_{eff}} P_p(L)\exp(-\alpha_p L)\right)} \quad (17)$$

Moreover, a semi-analytic solution (18) can be used to fit the simplified signal power profile: [32]

$$P_s(z) = \exp(-\alpha_s z) + b_2 \exp[-\alpha_2(L-z)] \quad (18)$$

In (18), the signal power profile is normalized to 1 mW (0

dBm) at $z=0$. The first term $\exp(-\alpha_s z)$ is the fiber attenuation term and the second term $b_2 \exp[-\alpha_2(L-z)]$ is the backward Raman amplification term, where $b_2$ and $\alpha_2$ are coefficients which can be obtained by simplifying (17):

$$b_2 \approx \frac{\exp(-\alpha_s L)\exp\left[\frac{g_R}{\alpha_p A_{eff}}P_p(L)\right]}{\exp\left(\frac{g_R}{\alpha_p A_{eff}}P_p(L)\exp(-\alpha_p L)\right)} \quad (19)$$

$$\alpha_2 \approx \frac{g_R}{A_{eff}}P_p(L) \quad (20)$$

Substituting (18), (19) and (20) into (10), the residual nonlinear noise power spectral density in systems with DRA is expressed as

$$I_{PMD} = \frac{3\gamma^2 I^3}{64\pi|\beta_2|}\left[1+\frac{b_2\alpha_s}{\alpha_2}\frac{1-\exp(\alpha_2 L)}{1-\exp(\alpha_s L)}\right]\left(\frac{1}{\alpha_s}+\frac{b_2^2}{\alpha_2}\right)$$

$$\cdot\left\{8N_s \ln\left(\frac{B}{B_0}\right) - \sum_{M=\frac{1}{2}}^{N_s-\frac{1}{2}}[3E_1(M)+E_2(M)]\right\} \quad (21)$$

And the ASE noise power is given by

$$P_{ASE}^{DRA} = N_s \cdot n_{sp}h\nu_0 g_R \Delta\nu_o \frac{G(L)}{A_{eff}}\int_0^L \frac{P_p(z)}{G(z)}dz \quad (22)$$

where $G(z)$ and $P_p(z)$ are Raman amplification gain and pump power at distance $z$ [30], [31].

## III. NUMERICAL VALIDATION OF THE PMD INFLUENCE ON IDEAL IC-NLC

In order to validate the theoretical analysis in the previous section and provide a quantitative assessment of the influence of non-deterministic PMD on IC-NLC, we conduct a single-channel numerical simulation. With PMD, the SNR of receiver signals becomes stochastic. In order to reasonably evaluate system performance, statistical distribution of SNR is first analyzed. We then use IDBP to completely eliminate deterministic intra-channel nonlinearity to obtain the optimal system performance and verify the theoretical analysis. Finally, the performance of IDBP is compared across various symbol rates.

### A. Simulation Setup

Fig. 2 depicts the simulation setup, and Table III lists the simulation parameters. In order to assess the influence of PMD on the effectiveness of IC-NLC, we select PMD parameters of 0.05 ps/km$^{1/2}$ and 0.1 ps/km$^{1/2}$ [35] and repeat the simulation multiple times to include different fiber polarization state realizations arising. This section presents the simulations of single-channel systems with varying symbol rates. At the transmitter, bit sequences are generated and mapped to symbol sequences. The modulation format is 16-ary quadrature amplitude modulation (16-QAM). Symbol rates of 100 GBaud, 200 GBaud and 300 GBaud are considered. The generated signal is pulse shaped by a root-raised cosine (RRC) filter whose roll-off factor is 0.02. The transmission link consists of multiple spans of SSMF. The fiber transmissions are simulated by SSFM. EDFA amplification and DRA amplification are both considered. The fiber loss is completely compensated for by an EDFA or a DRA in each span. ASE noise power of EDFA and DRA is calculated by (14) and (22), respectively. At the receiver, chromatic dispersion (CD) and fiber nonlinearity are compensated for by IDBP. Compared to low-complexity DBP, IDBP possesses a sufficiently fine propagation step size, enabling it to comprehensively compensate for deterministic intra-channel fiber nonlinearity. Fig. 3 presents the procedure of DBP/IDBP, taking an EDFA-amplified link as an example. DBP is based on SSFM. The dispersion and nonlinear effects in the real fiber are canceled by the dispersion acts and nonlinearity acts in the virtual fiber:

$$\widehat{D} = -\frac{j\beta_2}{2}\frac{\partial^2}{\partial t^2}+\frac{\alpha}{2} \quad (23)$$

$$\widehat{N} = -\frac{8}{9}j\gamma|E|^2 \quad (24)$$

where $\widehat{D}$ and $\widehat{N}$ are linear operator and nonlinear operator, respectively. Afterwards, other digital signal processing (DSP) algorithms such as linear equalization and carrier phase recovery are performed. Finally, SNR is calculated to evaluate the transmission performance.

### B. Simulation Results

In the SSFM-based simulations, the polarization effects and the interaction between PMD and nonlinearity are taken into account. Given the diverse fiber polarization state realizations arising from the influence of PMD, simulations are conducted with 75 PMD realizations for each launch power to accurately

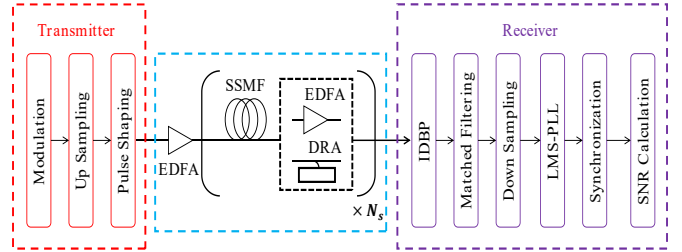

Fig. 2. Simulation setup of single-channel systems with EDFA and DRA. $N_s$: number of spans; IDBP: ideal digital back-propagation; LMS-PLL: least mean square phase-locked loop.

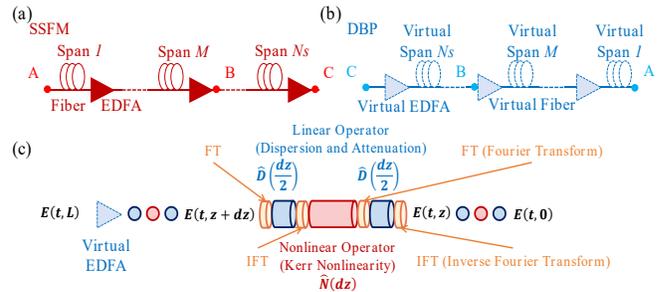

Fig. 3. (a) The schematic diagram of SSFM simulation link. (b) The schematic diagram of virtual link in DBP. (c) The procedure diagram of DBP compensation steps. EDFA-amplified link is taken as an example. $N_s$: number of spans; span M: $M^{th}$ span in the simulation systems; dz: DBP step-size.

TABLE III
SIMULATION PARAMETERS

| Parameters | Values |
|---|---|
| Symbol Rate (GBaud) | 100/200/300 |
| Number of Channel | 1 |
| Signal Center Wavelength (nm) | 1550 |
| Number of Spans | 10 |
| Span Length (km) | 80 |
| Attenuation Coefficient (dB/km) | 0.2 |
| Dispersion Coefficient (ps·nm$^{-1}$·km$^{-1}$) | 16.7 |
| Nonlinear Coefficient (W$^{-1}$·km$^{-1}$) | 1.3 |
| PMD Coefficient (ps/km$^{1/2}$) | 0.05/0.1 |
| Effective Area (um$^2$) | 80 |
| EDFA Gain (dB) | 16 |
| EDFA NF (dB) | 5 |
| DRA On-off Gain (dB) | 16 |
| DRA Pump Type | Backward |
| DRA Pump Wavelength (nm) | 1450 |

and comprehensively assess the influence of PMD.

The statistical distribution of SNR at the optimal launch power in systems with IDBP is illustrated in Fig. 4. SNRs under the influence of nonlinear signal-ASE interaction are represented by the black dashed curves. Average SNRs with both ASE and PMD are represented by solid curves. The discrepancy between black dashed curves and solid curves is highlighted to compare the impact of PMD in systems with different symbol rates. The SNR values under different PMD realizations are predominantly concentrated around the mean values. The probability of SNRs decreases as SNRs deviate further from the average SNRs. On the other hand, the average SNRs decrease as symbol rate increases. As the symbol rate increases from 100 GBaud to 300 GBaud, the discrepancy of the average SNR increases from 2.15 dB to 3.85 dB in EDFA-amplified links with a PMD parameter of 0.05 ps/km$^{1/2}$. In DRA-amplified links, the discrepancy of the average SNR with a PMD parameter of 0.05 ps/km$^{1/2}$ increases from 3.81 dB to 5.09 dB. This indicates that the influence of PMD is much stronger than ASE, and PMD becomes a significant limiting factor to the fiber channel capacity in ultra-high symbol rate systems. As PMD increases from 0.05 ps/km$^{1/2}$ to 0.1 ps/km$^{1/2}$, the average SNR decreases by 1.49 dB and 1.70 dB, respectively, in EDFA-amplified links and DRA-amplified links with a symbol rate of 300 GBaud.

Comparisons between numerical and theoretical results in EDFA-amplified links and DRA-amplified links are shown in Fig. 5 and Fig. 6, respectively. SNRs obtained through (13) are represented by solid lines, and average SNRs obtained from simulations are represented by diamonds. As the influence of nonlinear signal-ASE interaction is negligible [23], we compare the results of the closed-form expressions SSFM-based simulations directly. It is observed that the residual nonlinear noise power calculated by (10) and (21) well matches the simulation results. For systems with EDFA amplification, the small discrepancies between the simulation results and theoretical results can be attributed to the incomplete traversal of PMD realizations and the penalty introduced by DSP algorithms. For systems with DRA amplification, apart from the aforementioned reasons, the error between the fitted formula and the actual signal power profile can also introduce a certain degree of inaccuracy, especially in systems where the pump depletion cannot be neglected.

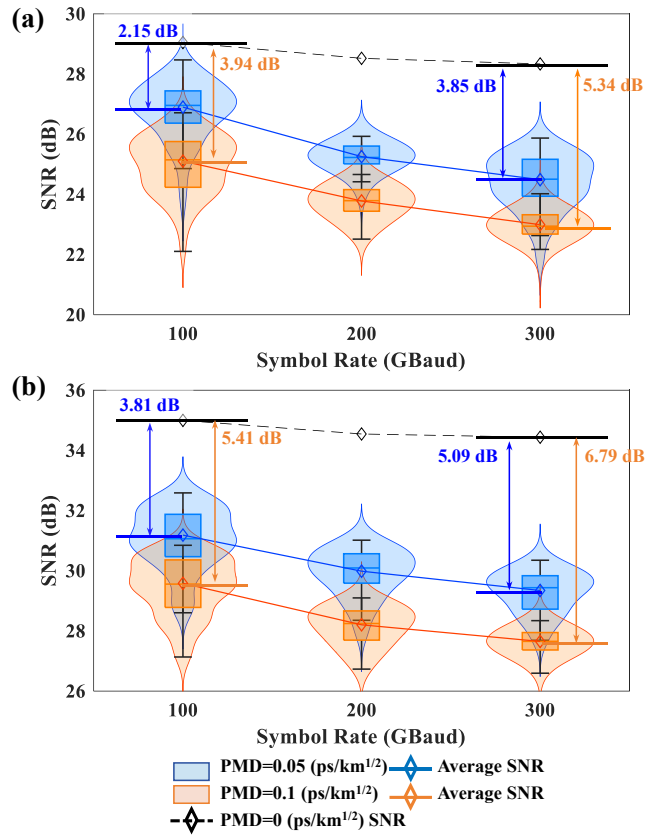

Fig. 4. Statistical distribution of SNR versus symbol rate for transmission over 10×80 km spans of SSMF in (a) EDFA-amplified links and (b) DRA-amplified links. For each combination of symbol rate and PMD parameter, 75 PMD realizations are simulated.

The performance of the EDFA-amplified links is first discussed. As the PMD parameter increases from 0.05 to 0.1 ps/km$^{1/2}$, the optimal launch power for IDBP-compensated system decreases by approximately 2 dB. In 300 GBaud systems, the peak SNR achieved by IDBP drops by 2.6 dB. Consequently, the residual nonlinear noise induced by PMD becomes more significant as the PMD parameter increases. Furthermore, the results from systems with different symbol rates are discussed to compare the impact of PMD on IDBP across these systems. Taking a PMD parameter of 0.05 ps/km$^{1/2}$ as an example, the peak SNR decreases from 27.29 dB to 25.65 dB, while the optimal launch power spectral density drops from 0.159 mW/GHz to 0.084 mW/GHz as the symbol rate increases from 100 GBaud to 300 GBaud. These simulation results indicate that the influence of PMD on ultra-high symbol rate systems is more significant. The SNRs at the optimal power for systems without IC-NLC are indicated by the black dashed curves. For a symbol rate of 300 GBaud, compared to systems without IC-NLC, the maximum SNR of systems with IDBP is improved by 6.51 dB and 3.91 dB for a PMD parameter of 0.05 and 0.1 ps/km$^{1/2}$, respectively. The results reveal that the magnitude of PMD has a significant impact on the effectiveness of IDBP. On the other hand, even with a relatively large PMD, there is still a large gain to achieve by compensating fiber nonlinearity. Note that the results are based on the average performance, and there are many fiber polarization state realizations where the system performance will be worse than these results.

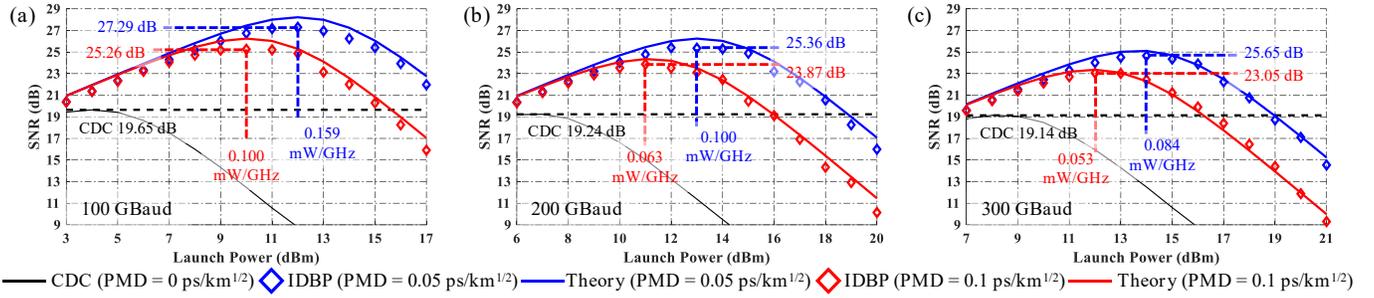

Fig. 5. SNR versus launch power in single channel EDFA-amplified links for (a) symbol rate of 100 GBaud, (b) symbol rate of 200 GBaud and (c) symbol rate if 300 GBaud. Black dashed lines are SNR without IC-NLC at optimal launch power. For each combination of symbol rate and PMD parameter, 75 PMD realizations are simulated.

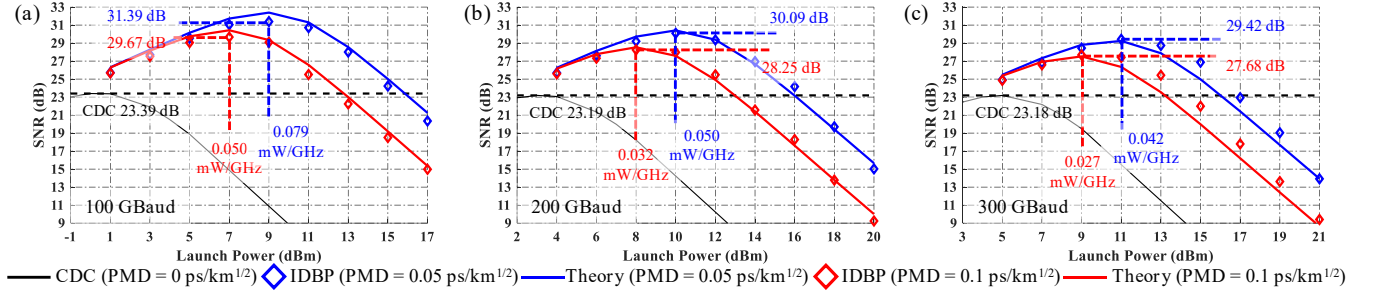

Fig. 6. SNR versus launch power in single channel DRA-amplified links for (a) symbol rate of 100 GBaud, (b) symbol rate of 200 GBaud and (c) symbol rate of 300 GBaud. Black dashed lines are SNR without IC-NLC at optimal launch power. For each combination of symbol rate and PMD parameter, 75 PMD realizations are simulated.

For the DRA-amplified links, the simulation results are similar to those of the EDFA-amplified links. As the symbol rate increases from 100 GBaud to 300 GBaud, the peak SNR decreases from 31.39 dB to 29.42 dB, with a PMD parameter of 0.05 ps/km$^{1/2}$. Meanwhile, the optimal launch power spectral density drops from 0.079 mW/GHz to 0.042 mW/GHz. As the symbol rate increases, the influence of PMD also becomes increasingly pronounced. For a symbol rate of 300 GBaud, the SNR gain with a PMD parameter of 0.05 and 0.1 ps/km$^{1/2}$ is 6.24 and 4.50 dB, respectively.

In this section, the influence of PMD in single-channel systems is simulated. The results demonstrate that PMD impairs the effectiveness of IDBP, particularly in ultra-high symbol rate systems. The analytical formulas of the residual nonlinear noise induced by PMD are consistent with simulations. So, these analytical formulas can be used to estimate performance for systems with different symbol rates and amplification schemes. In the next section, a more comprehensive exploration of IC-NLC in WDM systems is carried out to assess the performance of practical IC-NLC.

IV. SIMULATIONS OF PRACTICAL IC-NLC'S PERFORMANCE

In the previous sections, we have respectively investigated the SCI proportion and the influence of non-deterministic distributed PMD. These two factors are simultaneously taken into account in this section by simulations of C-band WDM systems. IDBP is first employed to evaluate the achievable gain of ideal IC-NLC in WDM systems with different symbol rates. We then use LDBP as an example to assess the potential gain of practical IC-NLC in WDM systems.

A. Simulation Setup

The simulation setup is depicted in Fig. 7, and Table IV lists the simulation parameters. In the transmitter, 16-QAM symbols are generated and then passed through a RRC filter with a roll-off factor of 0.02. After that, signals from different WDM channels are multiplexed. In order to evaluate the performance of practical systems, C-band (4 THz) WDM transmission simulations are carried out. The number of channels is 39, 19, and 13, corresponding to the symbol rate of 100 GBaud, 200 GBaud and 300 GBaud, respectively. The simulation transmission link consists of 10 SSMF spans with a span length of 80 km. An EDFA or DRA is used to compensate for the fiber loss in each span. We employ SSFM to simulate dispersion, fiber nonlinearity, and PMD within a bandwidth range of approximately 4 THz centered on a wavelength of 1550 nm.

In the receiver, matched filtering is first applied. Then CD and SCI are compensated for by LDBP or IDBP. For the LDBP algorithm, low-pass filtering (LPF) is performed in each nonlinearity compensation step to eliminate the high frequency components of the intensity [10]. LDBP has been proposed for several years and has been extensively studied, making it a relatively reliable algorithm. Its effectiveness has been demonstrated in various scenarios, including high-speed and long-haul transmission systems. In our simulation, we choose the Gaussian LPF and adjust its bandwidth to achieve optimal compensation performance. Various computational complexities of LDBP are considered for comparative analysis. After carrier phase recovery, SNR is calculated to evaluate the transmission performance. The SNR calculated by theoretical expressions is also taken into consideration for comparison.

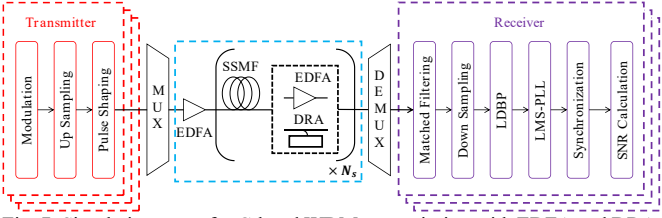

Fig. 7. Simulation setup for C-band WDM transmission with EDFA and DRA. $N_s$: number of spans; IDBP: ideal digital back-propagation; LMS-PLL: least mean square phase-locked loop; MUX: multiplexer; DEMUX: demultiplexer.

TABLE IV
SIMULATION PARAMETERS

| Parameters | Values |
| --- | --- |
| Symbol Rate (GBaud) | 100/200/300 |
| Channel Spacing (GHz) | 110/220/330 |
| Total Bandwidth | C-band (~4 THz) |
| Number of Channels | 39/19/13 |
| Signal Center Wavelength (nm) | 1550 |
| Number of Spans | 10 |
| Span Length (km) | 80 |
| Attenuation Coefficient (dB/km) | 0.2 |
| Dispersion Coefficient (ps·nm$^{-1}$·km$^{-1}$) | 16.7 |
| Nonlinear Coefficient (W$^{-1}$·km$^{-1}$) | 1.3 |
| PMD Coefficient (ps/km$^{1/2}$) | 0.05/0.1 |
| Effective Area (um$^2$) | 80 |
| EDFA Gain (dB) | 16 |
| EDFA NF (dB) | 5 |
| DRA On-off Gain (dB) | ~16 |
| DRA Pump Type | Backward |
| DRA Pump Wavelength (nm) | 1450 |

*B. Simulation Results*

The comparisons between the numerical and theoretical results in systems with a total bandwidth of 4 THz are shown in Fig. 8. The results indicate that the theoretical expressions maintain agreement with the SSFM results in 4 THz systems with different symbol rates. The results demonstrate that inter-channel nonlinearity does not compromise the accuracy of (10) and (21). Therefore, the interaction between PMD and inter-channel nonlinearity is independent of the interaction between PMD and intra-channel nonlinearity. As the symbol rate increases from 100 GBaud to 300 GBaud, the impact of PMD on IDBP performance becomes more significant. For systems with a symbol rate of 300 GBaud, as the PMD parameter increases from 0.05 ps/km$^{1/2}$ to 0.1 ps/km$^{1/2}$, the performance of IDBP decreases by 0.19 dB in EDFA/DRA-amplified links. Compared to the results in Fig. 5 and Fig. 6, the impact of PMD variations on IDBP effectiveness is relatively weaker in simulation systems with a total bandwidth of 4 THz. This phenomenon indicates that inter-channel nonlinearity noise accounts for a large proportion of the noise after IDBP compensation, thereby reducing the relative contribution of PMD-induced residual nonlinear noise. On the other hand, SNR in EDFA/DRA-amplified links with a PMD parameter of 0.05 ps/km$^{1/2}$ increases 0.63/0.82 dB as symbol rate increases from 100 GBaud to 300 GBaud, respectively. The simulation results indicate that the proportion of deterministic intra-channel nonlinearity noise, which can be compensated by IC-NLC, still increases with the increase of symbol rate. Therefore, IC-NLC has a high potential gain in ultra-high symbol rate systems.

Fig. 9 and Fig. 10 show the compensation performance of IDBP and LDBP in the EDFA-amplified links and DRA-amplified links, respectively. The simulation results indicate that the gain of IDBP increases with the increase of symbol rate in WDM systems. As the symbol rate increases from 100 GBaud to 300 GBaud, the SNR gain of IDBP increases from 0.65 dB to 1.36 dB for EDFA-amplified links, and from 1.02 dB to 1.76 dB for DRA-amplified links with a PMD parameter of 0.05 ps/km$^{1/2}$. On the other hand, compared to single-channel systems, the residual nonlinear noise after IDBP in WDM systems is primarily composed of inter-channel nonlinear effects rather than residual nonlinear noise induced by PMD. However, as the symbol rate increases, the performance gap of IDBP between systems with different PMD parameters becomes more pronounced. For the simulation systems with a symbol rate of 100 GBaud at the optimal launch power, the performance difference of IDBP is below 0.1 dB comparing PMD parameters of 0.05ps/km$^{1/2}$ and 0.1 ps/km$^{1/2}$. As the symbol rate increases to 300 GBaud, this performance gap widens to approximately 0.19 dB. Therefore, simulation results reveal that the influence of PMD on the performance of IDBP becomes non-negligible in ultra-high symbol rate systems.

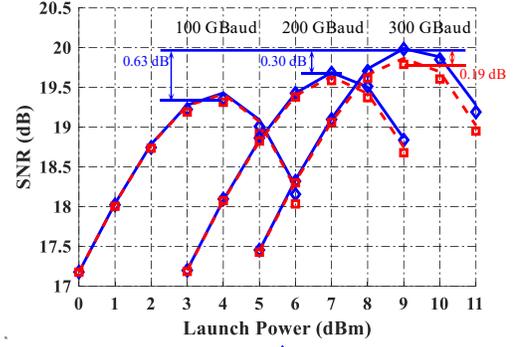

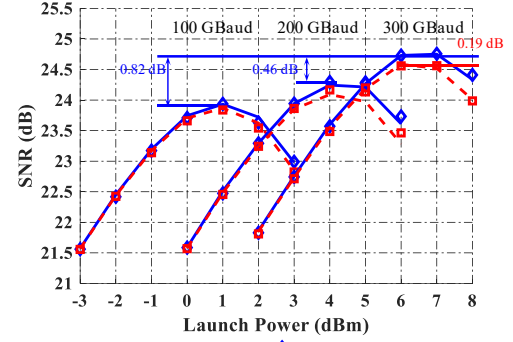

Fig. 8. SNR versus launch power in 4-THz IDBP compensated systems with (a) EDFA amplification and (b) DRA amplification. For each combination of symbol rate and PMD parameter, 5 PMD realizations are simulated.

Finally, the compensation performance of LDBP is discussed. The SNR gain of LDBP also increases with the increase of symbol rate. In the EDFA-amplified links with a PMD parameter of 0.05 ps/km$^{1/2}$, the SNR gain of the 20-stps LDBP increases from 0.53 dB to 0.87 dB as the symbol rate increases from 100 GBaud to 300 GBaud. However, the gains of 20-stps LDBP in 200 GBaud and 300 GBaud systems are quite similar, indicating a saturation in the performance of LDBP in ultra-high symbol rate systems. This is because the signal waveform changes very rapidly during transmission in ultra-high symbol rate systems. In systems with a symbol rate of 300 GBaud, LDBP struggles to accurately estimate high-frequency components of fiber nonlinearity noise. In the DRA-amplified links, since the signal is amplified distributively, the compensation performance of DBP is highly dependent on the symmetry of the signal power profile along the virtual fiber. Therefore, LDBP requires more steps to effectively compensate for SCI. The SNR gain of the 20-stps LDBP increases from 0.89 dB to 1.30 dB as the symbol rate increases from 100 GBaud to 300 GBaud. Fig. 9(a) and Fig. 10(a) depict the simulation results of LDBP in 100 GBaud systems, it is observed that the SNR curves almost overlap with each other as the number of steps per span increases. Due to the low proportion of SCI within the overall nonlinearity in low symbol rate systems, the potential gain of practical IC-NLC is limited. However, in the 300 GBaud systems, the gain of LDBP increases as the number of steps per span increases, as shown in Fig. 9(d) and Fig. 10(d). Therefore, the high SCI proportion creates a substantial potential gain for practical IC-NLC in high symbol rate systems. On the other hand, it is observed that PMD causes a small performance degradation in LDBP, especially in systems with a high symbol rate. In systems with a symbol rate $\geqslant 200$ GBaud and a PMD parameter of 0.1 ps/km$^{1/2}$, the LDBP performance degradation caused by PMD is approximately 0.1 dB. The results reveal that practical IC-NLC can achieve higher nonlinearity compensation performance in WDM systems with a higher symbol rate. However, IC-NLC needs to be specifically designed to reduce the complexity required to compensate for nonlinear effects in ultra-high symbol rate systems.

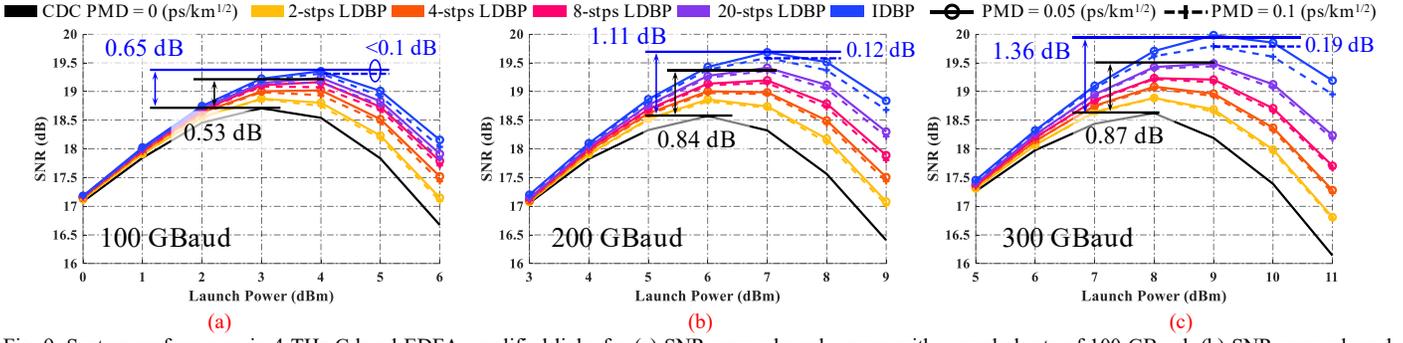

Fig. 9. System performance in 4-THz C band EDFA-amplified links for (a) SNR versus launch power with a symbol rate of 100 GBaud, (b) SNR versus launch power with a symbol rate of 200 GBaud and (d) SNR versus launch power with a symbol rate of 300 GBaud. For each combination of symbol rate and PMD parameter, 5 PMD realizations are simulated.

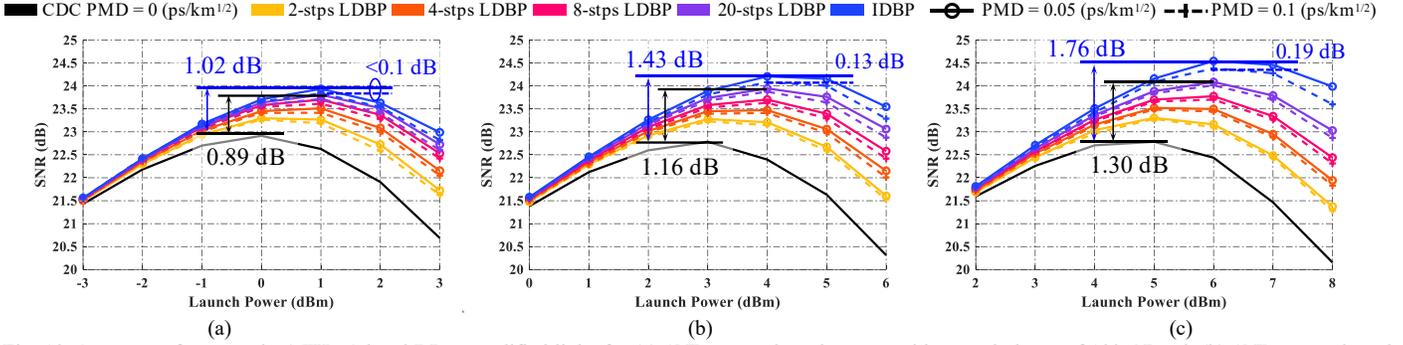

Fig. 10. System performance in 4-THz C band DRA-amplified links for (a) SNR versus launch power with a symbol rate of 100 GBaud, (b) SNR versus launch power with a symbol rate of 200 GBaud and (c) SNR versus launch power with a symbol rate of 300 GBaud. For each combination of symbol rate and PMD parameter, 5 PMD realizations are simulated.

## V. Future Research

As demonstrated by our evaluation, the deployment of IC-NLC is expected to achieve higher transmission capacity in WDM systems with a higher symbol rate. Nevertheless, further research is required before IC-NLC becomes a crucial technology for ultra-high symbol rate systems. The primary challenge is to achieve low-complexity IC-NLC while maintaining performance that meets the capacity requirement.

In ultra-high symbol rate systems, the intensity of signal waveform changes very rapidly during transmission, which requires increased computational complexity in IC-NLC algorithms for accurate estimation of the nonlinear effect correction. In addition, ultra-high symbol rate systems need to process substantial amounts of data in real time. These demands present significant challenges for designing low-complexity IC-NLC algorithms.

To reduce the computational complexity of IC-NLC, various low-complexity IC-NLC schemes have been widely investigated, such as LPF-assisted DBP [10], triplet-correlative PNC [14], inverse Volterra series transfer function [18] and so forth. These approaches also demonstrate potential in reducing the computational complexity of IC-NLC for ultra-high symbol rate systems. Since conventional fiber nonlinearity models typically rely on simplified assumptions derived from low symbol rate signal characteristics, fiber nonlinearity modeling of ultra-high symbol rate systems emerges as a promising direction for designing low-complexity IC-NLC. On the other hand, machine learning and other emerging technologies have made significant progress in reducing the computational complexity of IC-NLC [19], [20], demonstrating their potential for application in ultra-high symbol rate systems.

Last but not least, subcarrier multiplexing (SCM) systems have emerged as a promising solution for ultra-high symbol rate systems. Compared to single-carrier (SC) systems, SCM systems can achieve higher nonlinear tolerance by optimizing the symbol rate of each subcarrier [11], [12]. However, the increased subcarrier number in ultra-high symbol rate systems induces increased computational complexity of multi-subcarrier joint IC-NLC. Therefore, reducing the computational complexity of multi-subcarrier joint IC-NLC in ultra-high symbol rate SCM systems must also be addressed.

## VI. CONCLUSIONS

In this paper, we have quantitatively assessed the potential of IC-NLC in ultra-high symbol rate systems with a symbol rate of 200 GBaud and beyond. Our assessment provides useful guidance in designing IC-NLC algorithms for next generation ultra-high symbol rate systems.

First, we have provided a quantitative assessment of SCI proportion in systems with different symbol rates using the SSFM-based simulation. Taking a C-band SSMF system with EDFA amplification as an example, as the symbol rate increases from 100 GBaud to 300 GBaud, the SCI proportion increases from 45.4% to 66.5%. Therefore, IC-NLC exhibits a larger potential gain in systems with a higher symbol rate. Second, the influence of PMD is investigated. The power spectral density expressions of the residual nonlinear noise induced by PMD for single carrier WDM systems with EDFA and DRA are derived, which well match the results of SSFM-based simulations. Based on the theoretical expressions and SSFM-based simulations, the influence of non-deterministic PMD on the effectiveness of IC-NLC is investigated, showing that it becomes more significant as the symbol rate increases. In 300 GBaud systems with a PMD parameter of 0.05 ps/km$^{1/2}$, the influence of interaction between PMD and nonlinearity reduces the gain of IDBP by 3.85/5.09 dB in EDFA/DRA-amplified links, respectively. Finally, we assess the potential gain of the practical IC-NLC in C-band WDM systems. Simulation results indicate that as the symbol rate increases from 100 GBaud to 300 GBaud, the gain achieved by practical IC-NLC increases accordingly. For 800 km C-band SSMF transmission with a PMD parameter of 0.5 ps/km$^{1/2}$, compared to 100 GBaud systems, practical IC-NLC can achieve an additional 0.34/0.41 dB SNR gain in 300 GBaud systems with EDFA/DRA amplification, respectively. These quantitative results show that for 200 GBaud and beyond systems, there is a sizable gain to achieve by IC-NLC even with a large non-deterministic PMD. Our assessment offers guidance to both academia and industry to research and design fiber nonlinearity compensation.

## APPENDIX

The derivation process of (21) is shown in this section. First, we assume that pump depletion is negligible and first-order Raman amplification is chosen. (16) is simplified to

$$\pm \frac{dP_p^{\pm}(z)}{dz} = -\alpha_p P_p^{\pm}(z) \qquad (25)$$

The solution of (25) is

$$\begin{cases} P_p^+(z) = P_p^+(0)exp(-\alpha_p z) \\ P_p^-(z) = P_p^-(L)exp[-\alpha_p(L-z)] \end{cases} \qquad (26)$$

Then we combine (26) and (15):

$$\frac{dP_s(z)}{dz} = \frac{g_R(v_s, v_p)}{A_{eff}} P_s(z)[P_p^+(0)exp(-\alpha_p z) + P_p^-(L)exp(\alpha_p z)exp(-\alpha_p L)] - \alpha_s P_s(z) \qquad (27)$$

(27) conforms to the general formula of a first-order linear differential equation

$$y' + f(z)y = g(z) \qquad (28)$$

where $y = P_s(z)$, $f(z) = \frac{g_R(v_s, v_p)}{A_{eff}}[P_p^+(0)exp(-\alpha_p z) + P_p^-(L)exp(\alpha_p z)exp(-\alpha_p L)]$ and $g(z) = -\alpha_s$. According to the constant variation method, the general solution of (28) is

$$y = exp\left(\int f(z)dz\right)\left(\int g(z)exp\left(-\int f(z)dz\right)dz + \bar{C}\right) \qquad (29)$$

where $\bar{C}$ is a constant. It should be noted that $\int f(z)dz$ represents the original function of $f(z)$. According to (29), the solution of (27) is

$$P_s(z) = \frac{P_s(0)}{exp\left(-\frac{g_R(v_s, v_p)}{\alpha_p A_{eff}} P_p^+(0)\right) exp\left(\frac{g_R(v_s, v_p)}{\alpha_p A_{eff}} P_p^-(L)exp(-\alpha_p L)\right)}$$
$$\times exp\left[-\frac{g_R(v_s, v_p)}{\alpha_p A_{eff}} P_p^+(0)exp(-\alpha_p z)\right] \times$$
$$exp\left[\frac{g_R(v_s, v_p)}{\alpha_p A_{eff}} P_p^-(L)exp(-\alpha_p L)exp(\alpha_p z)\right] exp(-\alpha_s z) \qquad (30)$$

Considering backward DRA, $P_p^+(0) = 0$. (30) is simplified as

$$P_s(z) = \frac{P_s(0)}{exp\left(\frac{g_R(v_s, v_p)}{\alpha_p A_{eff}} P_p^-(L)exp(-\alpha_p L)\right)} \times$$
$$exp\left[\frac{g_R(v_s, v_p)}{\alpha_p A_{eff}} P_p^-(L)exp(-\alpha_p L)exp(\alpha_p z)\right] exp(-\alpha_s z) \qquad (31)$$

Replacing $b_2$ in (18) based on (31), we can get

$$\begin{cases} P_s(L) = exp(-\alpha_s L) + b_2 \\ P_s(0) = 1 + b_2 exp[-\alpha_2 L] \end{cases} \qquad (32)$$

Because (18) is the normalized signal power profile, we can obtain $P_s(L)$:

$$P_s(L) = \frac{exp\left[\frac{g_R(v_s,v_p)}{\alpha_p A_{eff}}P_p^-(L)\right]exp(-\alpha_s L)}{exp\left(\frac{g_R(v_s,v_p)}{\alpha_p A_{eff}}P_p^-(L)exp(-\alpha_p L)\right)} \quad (33)$$

As $exp(-\alpha_s L) \approx 0$, (32) can be transformed as

$$b_2 \approx \frac{exp(-\alpha_s L)exp\left[\frac{g_R}{\alpha_p A_{eff}}P_p^-(L)\right]}{exp\left(\frac{g_R}{\alpha_p A_{eff}}P_p^-(L)exp(-\alpha_p L)\right)} \quad (34)$$

Similarly, we consider the boundary conditions at $z = L$, $\alpha_2$ in (18) is

$$\alpha_2 \approx \frac{g_R}{A_{eff}}P_p(L) \quad (35)$$

In systems with lumped amplification, after digital back propagation, the residual FWM component can be expressed as Eq. (13) in [23], where

$$\int_0^L e^{-\alpha_s L}e^{-j\Delta\beta_{ijk}z}dz = \frac{1-e^{-\alpha_s L}e^{-j\Delta\beta_{ijk}L}}{j\Delta\beta_{ijk}+\alpha_s} \quad (36)$$

In systems with backward DRA, (36) becomes

$$\int_0^L [e^{-\alpha_s z} + b_2 e^{-\alpha_2 L}e^{\alpha_2 z}]e^{-j\Delta\beta_{ijk}z}dz =$$
$$\frac{1-e^{-\alpha_s L}e^{-j\Delta\beta_{ijk}L}}{j\Delta\beta_{ijk}+\alpha_s} + b_2 e^{-\alpha_2 L}\frac{1-e^{-\alpha_2 L}e^{-j\Delta\beta_{ijk}L}}{j\Delta\beta_{ijk}+\alpha_2} \quad (37)$$

Because the first item plays a dominant role in the first half of a fiber span, and the second item plays a dominant role in the second half of a fiber span. Therefore, these two items can be considered separately. According to symmetry of z, through the same derivation process as Eq. (17) in [23], the residual nonlinear noise power spectral density in systems with DRA is expressed as

$$I_{PMD} = \frac{3\gamma^2 I^3}{64\pi|\beta_2|}\left(\frac{1}{\alpha_s} + \frac{b_2^2}{\alpha_2}\right)$$
$$\cdot \left\{8N_s ln\left(\frac{B}{B_0}\right) - \sum_{M=\frac{1}{2}}^{N_s-\frac{1}{2}}[3E_1(M) + E_2(M)]\right\} \quad (38)$$

Considering that the assumption which two items in (37) can be considered separately, some residual nonlinear noise has been ignored. In systems with lumped amplification, effective length is

$$L_{eff} = \frac{1-exp(\alpha_s L)}{\alpha_s} \quad (39)$$

Based on the normalized signal power profile (18), effective length in systems with DRA is

$$L_{eff} = \frac{1-exp(\alpha_s L)}{\alpha_s} + b_2\frac{1-exp(\alpha_2 L)}{\alpha_2} \quad (40)$$

Thus, we obtain the correction coefficient

$$1 + \frac{b_2 \alpha_s}{\alpha_2}\frac{1-exp(\alpha_2 L)}{1-exp(\alpha_s L)} \quad (41)$$

Then we substitute correction coefficient into (38), and the residual nonlinear noise power spectral density in systems with DRA is expressed by the following equations:

$$I_{PMD} = \frac{3\gamma^2 I^3}{64\pi|\beta_2|}\left[1 + \frac{b_2 \alpha_s}{\alpha_2}\frac{1-exp(\alpha_2 L)}{1-exp(\alpha_s L)}\right]\left(\frac{1}{\alpha_s} + \frac{b_2^2}{\alpha_2}\right)$$
$$\cdot \left\{8N_s ln\left(\frac{B}{B_0}\right) - \sum_{M=\frac{1}{2}}^{N_s-\frac{1}{2}}[3E_1(M) + E_2(M)]\right\} \quad (42)$$